\documentclass[aps,prc,floatfix,showpacs,twocolumn, preprintnumbers]{revtex4}

\usepackage{bm}
\usepackage{color}
\usepackage{graphicx,epsfig}
\usepackage{amsfonts,amssymb,amsmath}
\usepackage{float}
\usepackage{xcolor}
\usepackage{color}
\usepackage{multirow}
\usepackage{physics}
\usepackage{subcaption}

\usepackage{dcolumn}     % Align table columns on decimal point

\newcommand{\nc}{\newcommand}

\newcommand{\beq}{\begin{equation}}
\newcommand{\eeq}{\end{equation}}
\nc{\bfx}{{\bf x}}
\nc{\bfy}{{\bf y}}
\nc{\bfz}{{\bf z}}
\nc{\bfxh}{{\bf \hat{x}}}
\nc{\bfyh}{{\bf \hat{y}}}
\nc{\bfzh}{{\bf \hat{z}}}
\nc{\bfj}{{\bf j}}
\nc{\bfr}{{\bf r}}
\nc{\bfR}{{\bf R}}
\nc{\bfk}{{\bf k}}
\nc{\bfq}{{\bf q}}
\nc{\bfp}{{\bf p}}
\nc{\bfv}{{\bf v}}
\nc{\bfs}{{\bf s}}
\nc{\bfA}{{\bf A}}
\nc{\bfJ}{{\bf J}}
\nc{\bfsg}{{\bm \sigma}}
\nc{\bfvh}{{\bf \hat{v}}}
\nc{\bfqh}{{\bf \hat{q}}}
\nc{\low}{\delta_{\rm Low}}

\nc{\rmel}[3]{\langle #1 || #2 || #3 \rangle}

\nc{\swap}{\rightleftharpoons}

\def\beq{\begin{equation}}
\def\eeq{\end{equation}}
\def\beqy{\begin{eqnarray}}
\def\eeqy{\end{eqnarray}}
\begin{document}

\title{Longitudinal form factors of $A\,\leq\,10$ nuclei in a chiral effective field theory approach}
\author{G.~B.~King$^{1}$}
\email{kingg@wustl.edu}
\author{G.~Chambers-Wall$^{1}$}
\email{chambers-wall@wustl.edu}
\author{A.~Gnech$^{2,3}$}
\email{agnech@odu.edu}
\author{S.~Pastore$^{1,4}$}
\email{saori@wustl.edu}
\author{M.~Piarulli$^{1,4}$}
\email{m.piarulli@wustl.edu}
\author{R~.B.~Wiringa$^5$}
\email{wiringa@anl.gov}

\affiliation{
$^1$\mbox{Department of Physics, Washington University in Saint Louis, Saint Louis, MO 63130, USA}\\
$^2$\mbox{Department  of  Physics,  Old  Dominion  University,  Norfolk,  VA  23529}\\
$^3$\mbox{Theory  Center,  Jefferson  Lab,  Newport  News,  VA  23610}\\
$^4$\mbox{McDonnell Center for the Space Sciences at Washington University in St. Louis, MO 63130, USA}\\
$^5$\mbox{Physics Division, Argonne National Laboratory, Argonne, IL 60439}\\
}

\begin{abstract}
In this work, we present the elastic electron scattering longitudinal form factors of $A\le 10$ nuclei computed in a variational Monte Carlo approach. We employ the Norfolk family of local chiral interactions and a consistent electromagnetic charge operator. Our calculations are compared both to data and past theoretical evaluations. This work represents, to our knowledge, the first accurate calculation of longitudinal form factors using many-body methods based on interacting nucleon degrees of freedom in the $7\,\leq\,A\,\leq\,10$ mass range. Finally, we identify $^9$Be and $^{10}$B as candidate targets for renewed experimental interest, as they exhibit the potential to provide more stringent constraints on the theoretical models. 
\end{abstract}

\maketitle

\section{Introduction}
\label{sec:intro}

In recent years, great advances have been made toward describing the nucleus as a collection of nucleons interacting amongst themselves via effective two- and three-nucleon forces~\cite{Hergert:2020bxy,Coraggio:2021jvf}. A major catalyst of this progress was the introduction of  chiral effective field theory ($\chi$EFT)~\cite{Weinberg:1978kz}, which has allowed for the development of many-nucleon interactions~\cite{Weinberg:1990rz,vanKolck:1994yi,Epelbaum:2008ga,Machleidt:2017vls,Piarulli:2022hml} and current operators~\cite{Park:1995pn,Walzl:2001vb,Phillips:2003jz,Phillips:2006im,Kolling:2009iq,Pastore:2011ip,Pastore:2008ui,Pastore:2009is,Piarulli:2012bn,Kolling:2011mt,Krebs:2016rqz,Krebs:2019aka} rooted in the underlying theory of quantum chromodynamics (QCD)~\cite{Gross:2022hyw}. Since the $\chi$EFT approach admits an expansion in the low-energy scale of nuclear physics relative to the high-energy scale of QCD, it is systematically improvable. This framework also enables the inclusion of relevant many-body effects up to  a specified order and, in principle, allows for an estimation of the impact of any missing sub-leading contributions.

Validating $\chi$EFT approaches with electromagnetic observables has several advantages, mainly due to the abundance of high-quality experimental data. Additionally, because the electron provides a relatively clean probe that minimally perturbs the nucleus, it allows for straightforward analyses of the measured cross sections~\cite{DeForest:1966ycn,Donnelly:1984rg}. The nuclear charge density extracted from electron scattering is a powerful tool to understand how nuclear dynamics impact ground state properties of the system~\cite{Piarulli:2022ulk}; however, direct comparisons to experimental charge form factors minimize the model dependence in the experimentally extracted quantities. Moreover, these data cover a broad range of momentum transferred to the nucleus, providing insights into where the $\chi$EFT expansion breaks down. 

Recent studies including the present authors' showed that an excellent agreement with measured magnetic moments and form factors in $3\,\leq\, A \,\leq\,10$ systems is achieved when using Norfolk models and associated many-nucleon electromagnetic current operators~\cite{Chambers-Wall:2024uhq,Chambers-Wall:2024fha}. The Norfolk models of two- and three-nucleon interactions, which correlate nucleons in pairs and triplets, are derived within the $\chi$EFT framework. This framework incorporates pions, nucleons and $\Delta$'s resulting in long- and intermediate-range interactions of one- and two-pion range, as well as short-range contributions represented by contact terms. In Refs.~\cite{Chambers-Wall:2024uhq,Chambers-Wall:2024fha}, we adopted electromagnetic one- and two-body vector  currents derived within the same $\chi$EFT framework as the Norfolk interactions. An accurate description of the data is achieved when accouting for two-nucleon currents. Most notably, two-body currents of one-pion range appear already at next-to-leading order (NLO) in the chiral expansion, providing significant corrections to the computed magnetic observables.   

In this work, we continue our investigations analyzing the longitudinal form factors from a $\chi$EFT perspective in nuclei within the same $3\,\leq\, A \,\leq\,10$ mass range investigated in Refs.~\cite{Chambers-Wall:2024uhq,Chambers-Wall:2024fha}. Specifically, we adopt the variational Monte Carlo (VMC) method to solve the many-body Schr\"odinger equation with Norfolk two- and three-nucleon interactions. The many-nucleon charge operator~\cite{Walzl:2001vb,Phillips:2003jz,Phillips:2006im,Kolling:2009iq,Pastore:2011ip} presents a very different expansion when compared to the  electromagnetic vector current operator. In fact, two-nucleon charge operators appear first at N3LO in the chiral expansion and consist of one-pion range operators. These are proportional to $1/m_N$, where $m_N$ is the nucleon mass, and may be viewed as relativistic/kinematic corrections. 

With this study, we aim at setting the foundation for a systematic analysis of electron scattering data within a $\chi$EFT framework, with the goal of validating the theoretical model of the nucleus. Using the Norfolk model, we are able to assess the impact of different fitting procedures for the two-body interaction on the longitudinal form factor and to investigate the role of two-body currents in its description. By calculating higher-order multipole contributions and looking at their behavior over a range of kinematics, we are able to see the imprint of nuclear structure and shape on predictions of electromagnetic observables. Finally, this work represents the next step toward an evaluation of elastic electron scattering cross sections using highly accurate many-body methods and nuclear interactions and electromagnetic currents from $\chi$EFT. This, in turn, can be  directly compared to experimentally measured cross sections, therefore avoiding the kinematic-dependent Rosenbluth separation procedure.

The remainder of the article is structured as follows: In Section~\ref{sec:theory}, we provide the definition of the longitudinal form factor and introduce the nuclear model adopted in the calculations. Finally, we present our results in Section~\ref{sec:results} and concluding remarks in Section~\ref{sec:conclusions}. 

\section{Theory}
\label{sec:theory}

In this section, we present the decomposition of the longitudinal form factor in multipoles, and express them in terms of nuclear matrix elements of the electromagnetic charge operator. The variational Monte Carlo method used to compute the matrix elements, along with the Norfolk interactions, have been most recently summarized in Ref.~\cite{Chambers-Wall:2024uhq}. Here, we briefly outline the key features relevant to the discussion of the results, with emphasis placed on the description of the $\chi$EFT charge operators adopted in the present work. 

\subsection{Multipole decomposition of $F_L(q)$}
\label{sec:multi}

The longitudinal form factor $F_L(q)$ is defined as~\cite{Walecka:1995},
\begin{equation}
F_L^2(q) = \frac{1}{2J+1} \sum_{l\geq0} |\rmel{J}{C_l(q)}{J}|^2\, ,\label{eq:fl.def}
\end{equation}
where $q$ is the magnitude of the three-momentum transferred to the nucleus, $C_l(q)$ is the $l^{\rm th}$ mulitpole excited by the scattering, and the double lines indicate a reduced matrix element of the multipole calculated in between nuclear states $|J\rangle$ of total angular momentum of $J$. Due to parity conservation, only even $l$ multipoles  with $0\leq l \leq 2J$ will contribute to the sum in Eq.~(\ref{eq:fl.def}). Thus, for integer $J$, there are $J+1$ multipoles that are excited by the scattering and for half-integer $J$, there are $J + 1/2$ multipoles. These are defined in terms of the charge operator, $\rho$, as~\cite{Carlson:1997qn}
\begin{eqnarray}\nonumber
\mel{JM}{\rho(q,\theta)}{JM} &=& (-1)^{J-M} \sum_l \sqrt{4\pi} (-i)^l P_l(\cos\theta)\\
&& \hspace{2cm} \times c^{M}_{JJl} C^J_l(q)\, , \label{eq:ch.ft}
\end{eqnarray}
where $\theta$ denotes the azimuthal angle formed by $\bfq$ with respect to the spin quantization axis taken to be along $\bfzh$, $P_l(\cos\theta)$ is the $l^{\rm th}$ Legendre polynomial, $C^J_l(q) = \rmel{J}{C_l(q)}{J}$, and the Clebsch-Gordan coefficient is defined as
\begin{equation}
c^M_{JJl} = \langle JM;JM | l0 \rangle\, .
\end{equation}
It is convenient to elect $M=J$ so that, 
\begin{equation}
\mel{JJ}{\rho(q,\theta)}{JJ} = \sum_l \sqrt{4\pi} (-i)^l P_l(\cos\theta) c^{J}_{JJl} C^J_l(q)\, . \label{eq:rho}
\end{equation}
To obtain the multipoles, one need only compute the matrix element in Eq.~(\ref{eq:rho}) for as many independent choices of $\theta$ as there are allowed multipoles, and then solve the resultant system of equations. In Appendix~\ref{sec:c.of.q}, we list the explicit expressions of the various $C_l(q)$ for all values of $J$ considered.

It is worth noting that all of the experimental data to which we compare normalize $F^2_L(0)$ to $1$ so that it represents the deviation from scattering on a point particle with charge $Z$. To appropriately normalize the standard definition of Eq.~(\ref{eq:fl.def}), we consider the behavior of the charge multipoles for $q=0$. We note that $\mel{JJ}{\rho(0,\theta)}{JJ}=Z$ for all $\theta$, $C_{l>0}(0)=0$~\cite{Walecka:1995}, and $c^J_{JJ0} = (2J+1)^{-1/2}$. Then, from Eq.~(\ref{eq:ch.ft}), it follows that,
\begin{equation}
C_0(0) = \sqrt{\frac{2J+1}{4\pi}}Z\,     
\end{equation}
and therefore,
\begin{equation}
F_L^2(0) = \frac{Z^2}{4\pi}\, .    
\end{equation}
To make connection with experimental data, we plot the form factor re-scaled as
\begin{equation}
F_L^2(q) \to \frac{4\pi}{Z^2} F_L^2(q)\, .   
\end{equation}

\subsection{Norfolk interactions}
\label{sec:vmc}

We perform the many-body calculations of the nuclear matrix elements using the variational Monte Carlo (VMC) method~\cite{Wiringa:1991kp,Pudliner:1997ck}. This technique retains the complexity of many-nucleon interactions and currents and has been extensively reviewed~\cite{Carlson:2014vla,Gandolfi:2020pbj}. When used with accurate many-nucleon Hamiltonians, quantum Monte Carlo (QMC) methods explain the data of several low-energy observables, including binding energies and spectra~\cite{Piarulli:2017dwd,Lonardoni:2018nob}, 
electromagnetic moments and transitions~\cite{Pastore:2012rp,Pastore:2014oda,Piarulli:2012bn,Gnech:2022vwr,Chambers-Wall:2024uhq,Chambers-Wall:2024fha}, electroweak responses~\cite{Lovato:2015qka,Lovato:2016gkq,Pastore:2019urn,Lovato:2017cux,Andreoli:2021cxo}, and weak observables~\cite{Pastore:2017uwc,King:2020wmp,King:2021jdb,King:2022zkz}.
 Key features relevant to this work have been summarized in our recent paper on the magnetic structure of nuclei~\cite{Chambers-Wall:2024uhq}, which forms the basis of this study. We refer the reader to that paper and the recent review of electroweak structure calculations using quantum Monte Carlo methods~\cite{King:2024zbv} for more details.

The many-body nuclear Hamiltonian, $H$, adopted in this work has the form
\begin{equation}
H = \sum_i T_i + \sum_{i<j} v_{ij} + \sum_{i<j<k} V_{ijk} + \ldots\, .   
\end{equation}
where $T_i$ is the one-body kinetic term. For the two- and three-nucleon interactions,  $v_{ij}$ and $V_{ijk}$, respectively, we use the Norfolk models of two- and three-body nuclear forces, collective referred to as the NV2+3 interactions~\cite{Piarulli:2014bda,Piarulli:2016vel,Piarulli:2017dwd,Baroni:2018fdn}. These local potential models are derived using a $\chi$EFT framework with nucleons, pions, and $\Delta$-isobars as degrees of freedom. The two-body force (NV2) consists of contact terms characterized by a  short-range regulator, $R_S$, associated with the Gaussian smearing of the $\delta$-functions in coordinate space, as well as intermediate- and long-range components of one- and two-pion range. The latter present singularities at short-distances that are removed using a regularization function with cutoff $R_L$. There are four classes of models for the NV2. Model classes denoted by I (II) are fit to $N\!N$ scattering data up to 125 (200) MeV with a $\chi^2/{\rm datum}\approx 1.1~(1.4)$. For model classes denoted by ``a" (``b"), the regulators used in the NV2 potential are $[R_L,R_S]=[1.2~{\rm fm},0.8~{\rm fm}]$ ($[R_L,R_S]=[1.0~{\rm fm},0.7~{\rm fm}]$). We will refer to the former and the latter choices as ``soft" and ``hard" interactions, respectively.  

In addition to the NV2, the Norfolk models incorporate a three-body force (NV3)~\cite{vanKolck:1994yi,Epelbaum:2002vt} that requires the determination of two additional LECs; namely,  $c_D$ and $c_E$. These are fit either to the trinucleon ground state energy and $nd$ doublet scattering length~\cite{Piarulli:2017dwd} or to the trinucleon ground state energy and Gamow-Teller matrix element for $\beta$-decay~\cite{Baroni:2018fdn}. The latter fitting procedure corresponds to the model classes denoted with a star.

\begin{figure}
\begin{center}
    \includegraphics[width=3.3in]{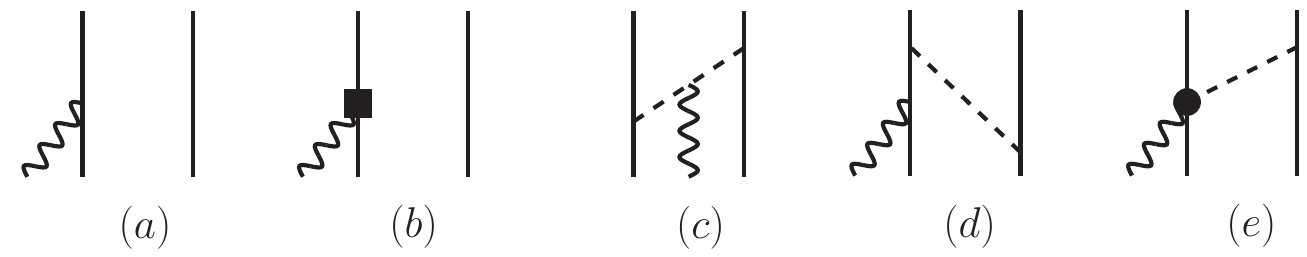}
\end{center}
\caption{Diagrams  illustrating  the  contributions  to  the  electromagnetic  charge  at  N3LO adopted in this work.  Specifically, panel (a), panel (b), panels (c)-(e) denote contributions at LO, N2LO, and N3LO, respectively. Nucleons,  pions,  and  external  fields  are  denoted  by  solid,  dashed,  and  wavy  lines, respectively. 
The square in panel (b) represents the $(Q/m_N)^2$, or $(v/c)^2$, relativistic
correction to the LO one-body charge operator, whereas the solid circle in panel (e) is associated
with a sub-leading $\gamma \pi N$ charge coupling. Figure from Ref.~\cite{Pastore:2011ip}. }
\label{fig:chargen3lo}
\end{figure}

\subsection{Charge operator to N3LO in chiral EFT}
\label{sec:charge.op}

The electromagnetic charge operator, ${\rho}$, is expressed as an expansion in many-body terms as
\begin{eqnarray}
{\rho} &=&  \sum_i {\rho}_i({\bf q}) + \sum_{i<j} {\rho}_{ij}({\bf q}) +\dots  \ ,
\end{eqnarray}
where ${\bf q}$ is the momentum transferred to the nucleus. The electromagnetic charge operator has been extensively studied within several implementations of $\chi$EFT~\cite{Walzl:2001vb,Phillips:2003jz,Phillips:2006im,Kolling:2009iq,Pastore:2011ip}.
In this work, we adopt the operators derived in Ref.~\cite{Pastore:2011ip}, and consider contributions up to N3LO in the chiral expansion. These are displayed in Fig.~\ref{fig:chargen3lo}, while their expressions in momentum space are provided in Refs.~\cite{Pastore:2011ip,Piarulli:2012bn}.

\begin{figure*}
\includegraphics[width=0.95\textwidth]{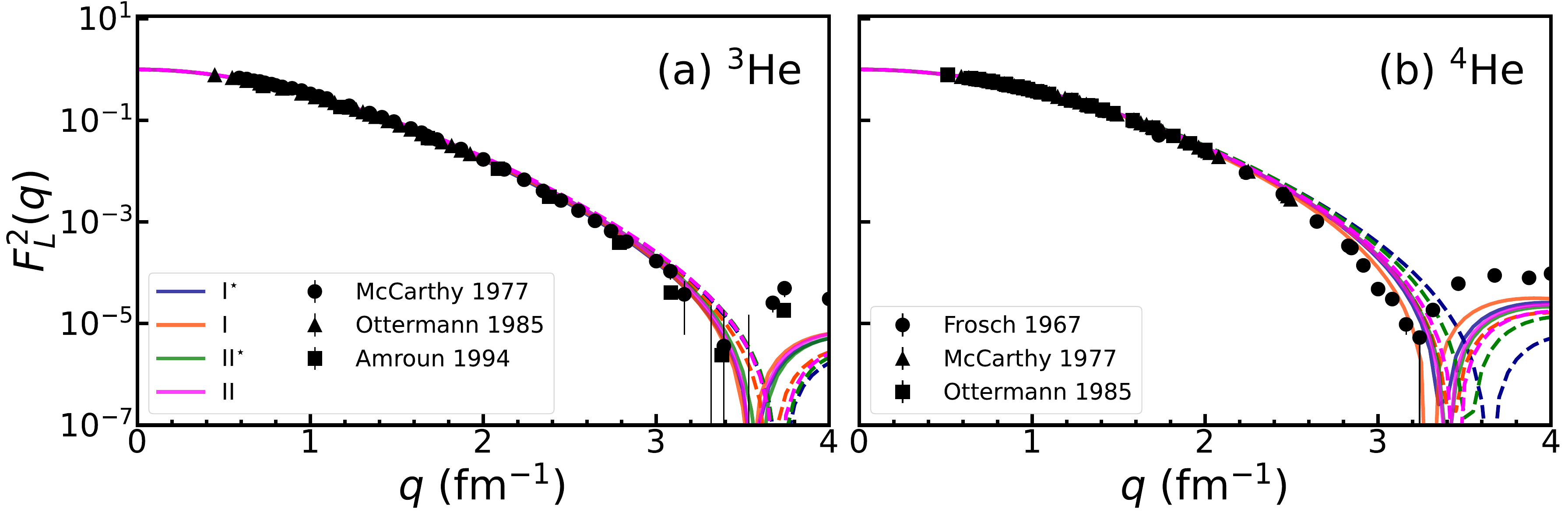}
    \caption{VMC calculations of the (a) $^3$He and (b) $^4$He longitudinal form factor using various NV2+3 model classes compared with experimental data (black symbols). The calculations using models I$^{\star}$ (blue lines), I (orange lines), II$^{\star}$ (green lines), and II (magenta lines) are shown with solid lines for the ``hard" cutoffs of model class b and with dashed lines for the ``soft" cutoffs of model class a.} 
    \label{fig:all.model.he}
\end{figure*}

One-body operators appear at LO and N2LO, and are represented by the disconnected diagrams displayed in panels (a) and (b) of Fig.~\ref{fig:chargen3lo}, respectively. The LO one-body operator, $\rho^{\rm LO}$, is derived from the non-relativistic reduction of the covariant single-nucleon current; that is, by expanding the operator in powers of $p_i/m_N\sim Q/m_N$, where $p_i$ and $m_N$ are the momentum and mass of a nucleon, respectively, and $Q$ denotes a small momentum characteristic of the low-energy regime of $\chi$EFT.  Evaluation of this disconnected diagram leads to the standard non-relativistic one-body charge in the impulse approximation~\cite{Carlson:1997qn}
\begin{equation}
\label{eq:lo}
\rho^{\rm LO} =  \epsilon_i(q^2) \,{\rm e}^{i{\bf q}\cdot{\bf r}_i} \ . 
\end{equation}
In the equation above
\begin{equation}
\epsilon_{i}(q^2) = \frac{G_E^S(q^2)+G_E^V(q^2)\, \tau_{i,z}}{2}\ , 
\label{eq:chargelo}
\end{equation}
where $G^{S/V}_E$ denote the isoscalar/isovector
combinations of the proton and neutron electric ($E$) 
form factors, normalized as $G^S_E(0)=G^V_E(0)=1$. 
%Note that Eq.~(\ref{eq:chargelo}) reduces to the proton projection operator in the limit of $q\rightarrow 0$.

The one-body operator at N2LO accounts for relativistic corrections of order $(Q/m_N)^2$ to the LO one-body operator. Evalulation of the disconnected diagram of panel (b) leads to the standard Darwin-Foldy and spin-orbit relativistic corrections~\cite{Pastore:2011ip,Carlson:1997qn}.

Two-body charge operators appear at N3LO. Specifically, while the diagrams of one-pion range displayed in panels (c) and (d) vanish in the static limit of $m_N\rightarrow\infty$, they contribute at N3LO once recoil corrections are accounted for~\cite{Pastore:2011ip}. The last N3LO contribution illustrated in panel (e) originates from a sub-leading $\gamma \pi N$ coupling appearing in the chiral Lagrangian~\cite{Pastore:2011ip}. This last term was first introduced within the $\chi$EFT framework by Phillips in Refs.~\cite{Phillips:2003jz,Phillips:2006im}. The need for such a contribution was long-known, having been derived two decades prior in the analysis of the non-relativistic virtual-pion photoproduction amplitude~\cite{Riska:1989bh}. 

A few comments are now in order. First, note that there are no contributions at NLO. Two-body terms are highly suppressed with respect to the LO contribution, as they appear for the first time at N3LO. From the expressions given in Ref.~\cite{Pastore:2011ip}, we see that two-body currents are proportional to $1/m_N$, and can therefore be considered as small kinematic corrections that vanish in the static limit. Moreover, there are no unknown Low Energy Constants entering the charge operator. Based on these last considerations, and with the goal of  streamlining the computations in $A>3$ systems, we have disregarded N4LO corrections in the calculations. The N4LO one-loop corrections of two-pion range have been derived in Ref.~\cite{PhysRevC.80.045502} by K\"olling {\it et al.} using the unitary transformation method, and subsequently in Ref.~\cite{Pastore:2011ip}, using time-ordered perturbation theory beyond the static limit, as it was done for the tree-level operators of Fig.~\ref{fig:chargen3lo} described in this section. This choice is also motivated by the results obtained in Ref.~\cite{Piarulli:2012bn} where the authors implemented the N4LO corrections to study the charge form factors of the deuteron and the trinucleon systems. It was found that they provide, on average, a correction $\lesssim 0.4$ \%  for values of momentum transfer in the range of $[0,5]$ fm$^{-1}$.

\section{Results}
\label{sec:results}

\begin{figure*}
\includegraphics[width=0.95\textwidth]{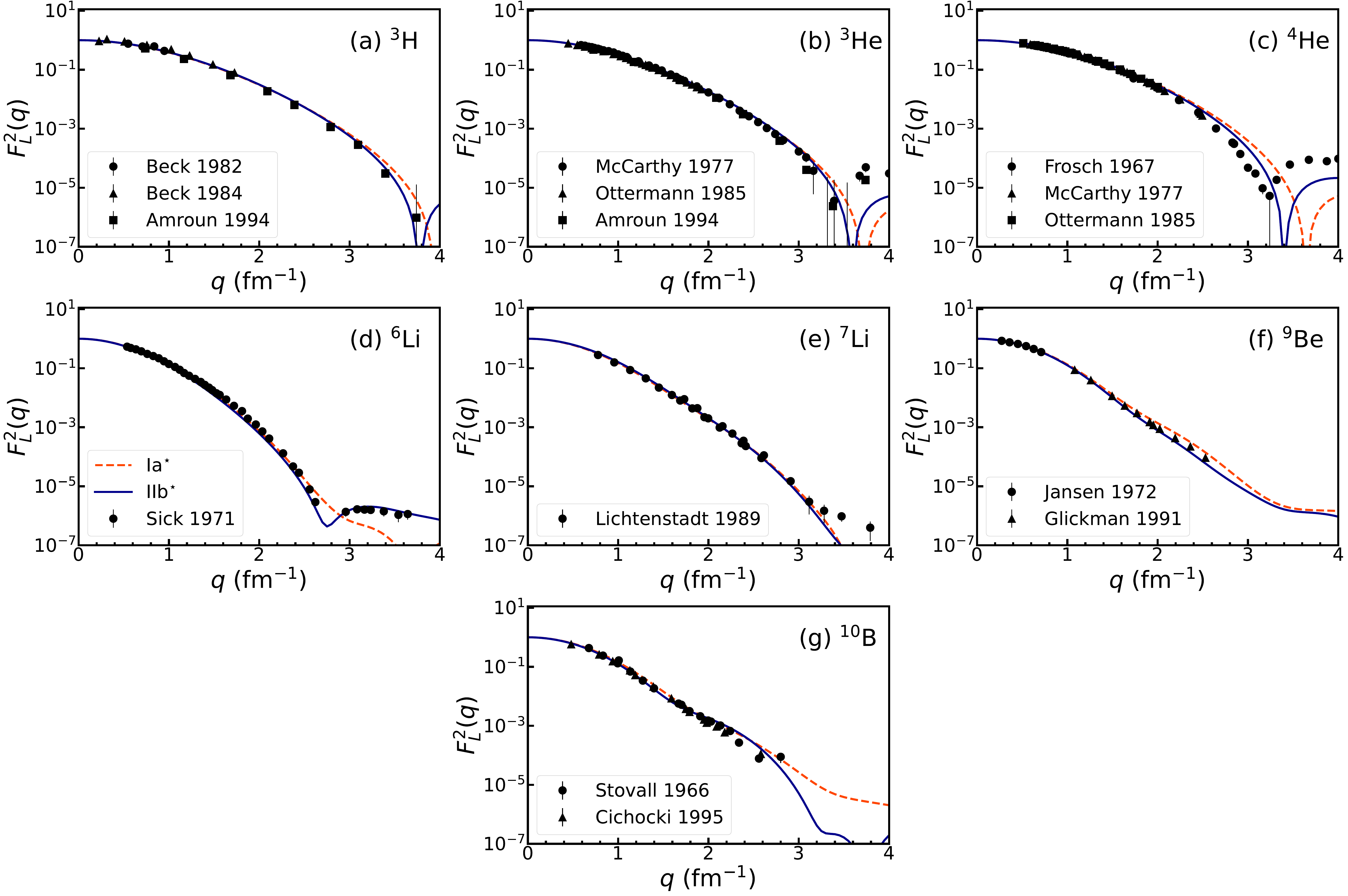}
    \caption{VMC calculations of the longitudinal form factors of various isotopes using models NV2+3-Ia$^{\star}$ (orange dashed line) and NV2+3-IIb$^{\star}$ (blue solid line) compared with experimental data (black symbols).} 
\label{fig:model.comp}
\end{figure*}

\begin{figure*}
\includegraphics[width=0.95\textwidth]{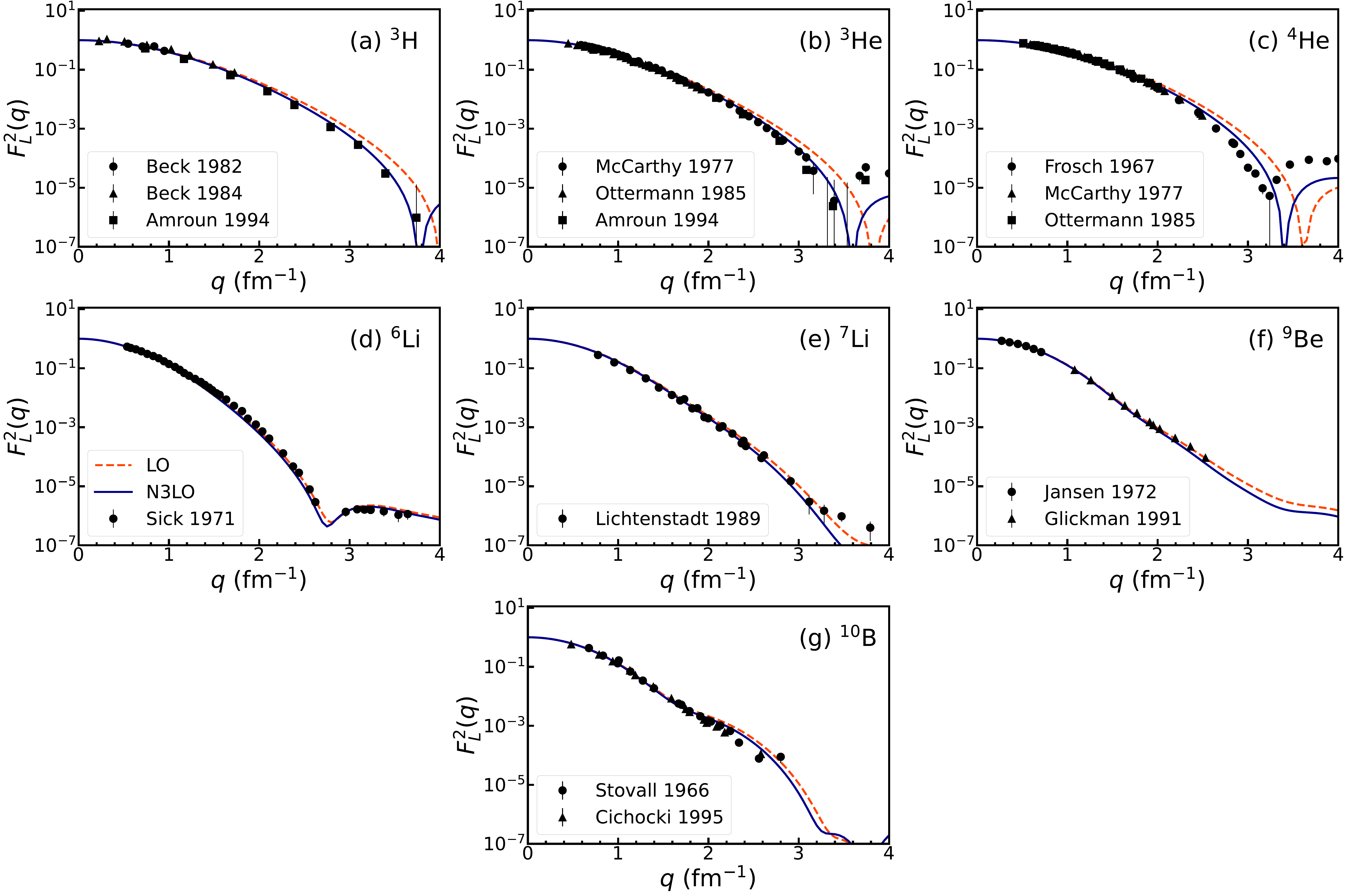}
    \caption{VMC calculations of the longitudinal form factor for various nuclei using model NV2+3-IIb$^{\star}$ performed with the LO charge operator (orange dashed line) and the charge operator with terms through N3LO (solid blue line) compared with data (black symbols).} 
\label{fig:two.body}
\end{figure*}

\begin{figure*}
\includegraphics[width=0.95\textwidth]{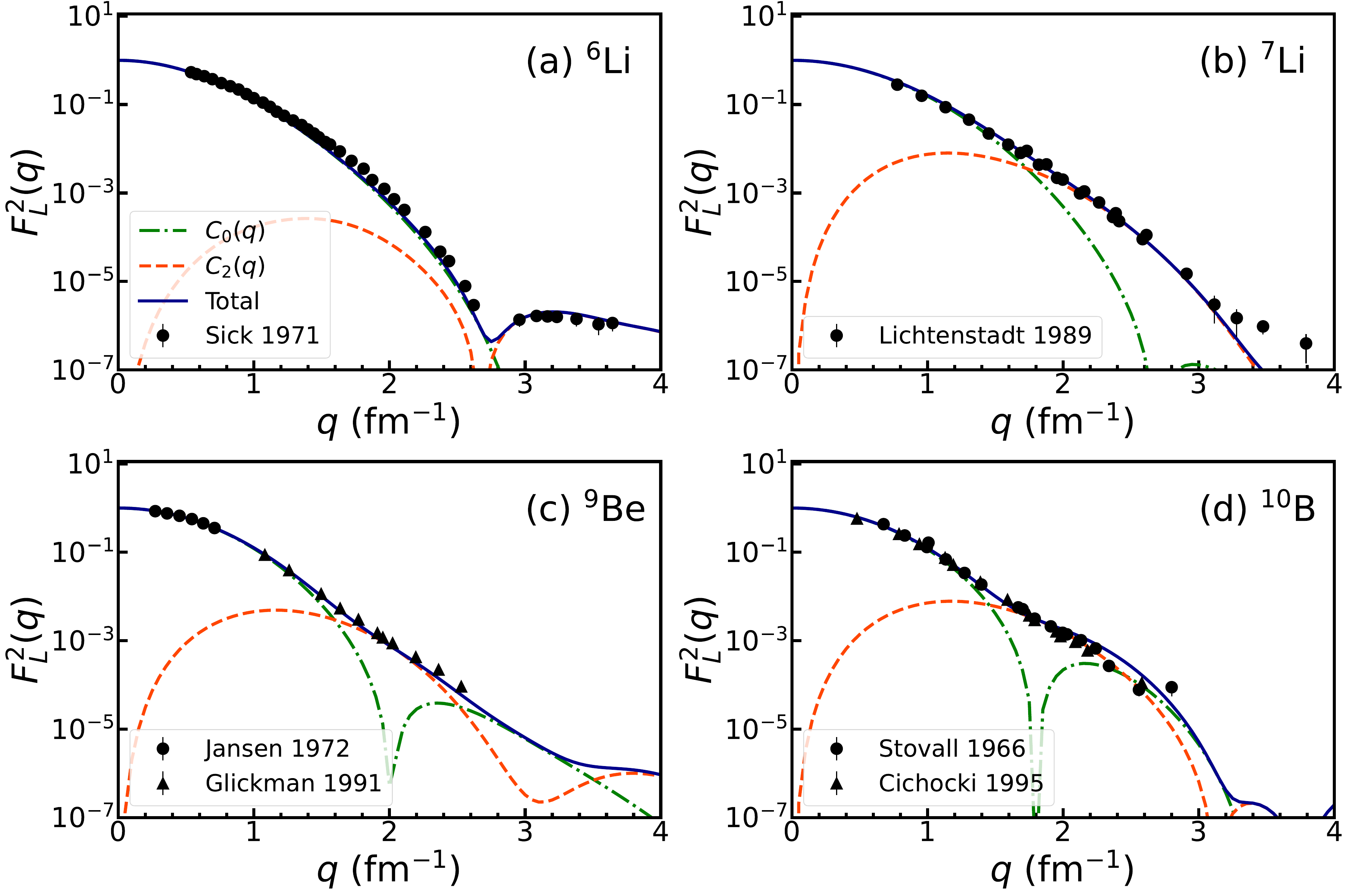}
    \caption{VMC calculations of the longitudinal form factor for various nuclei using model NV2+3-IIb$^{\star}$ and the charge operator through N3LO (solid blue line) compared to data (black symbols). The calculations are decomposed into the $C_0$ (green dot dashed line) and $C_2$ (orange dashed lines) multipole contributions.}  
\label{fig:multipole.119}
\end{figure*}

\begin{figure*}
\includegraphics[width=0.95\textwidth]{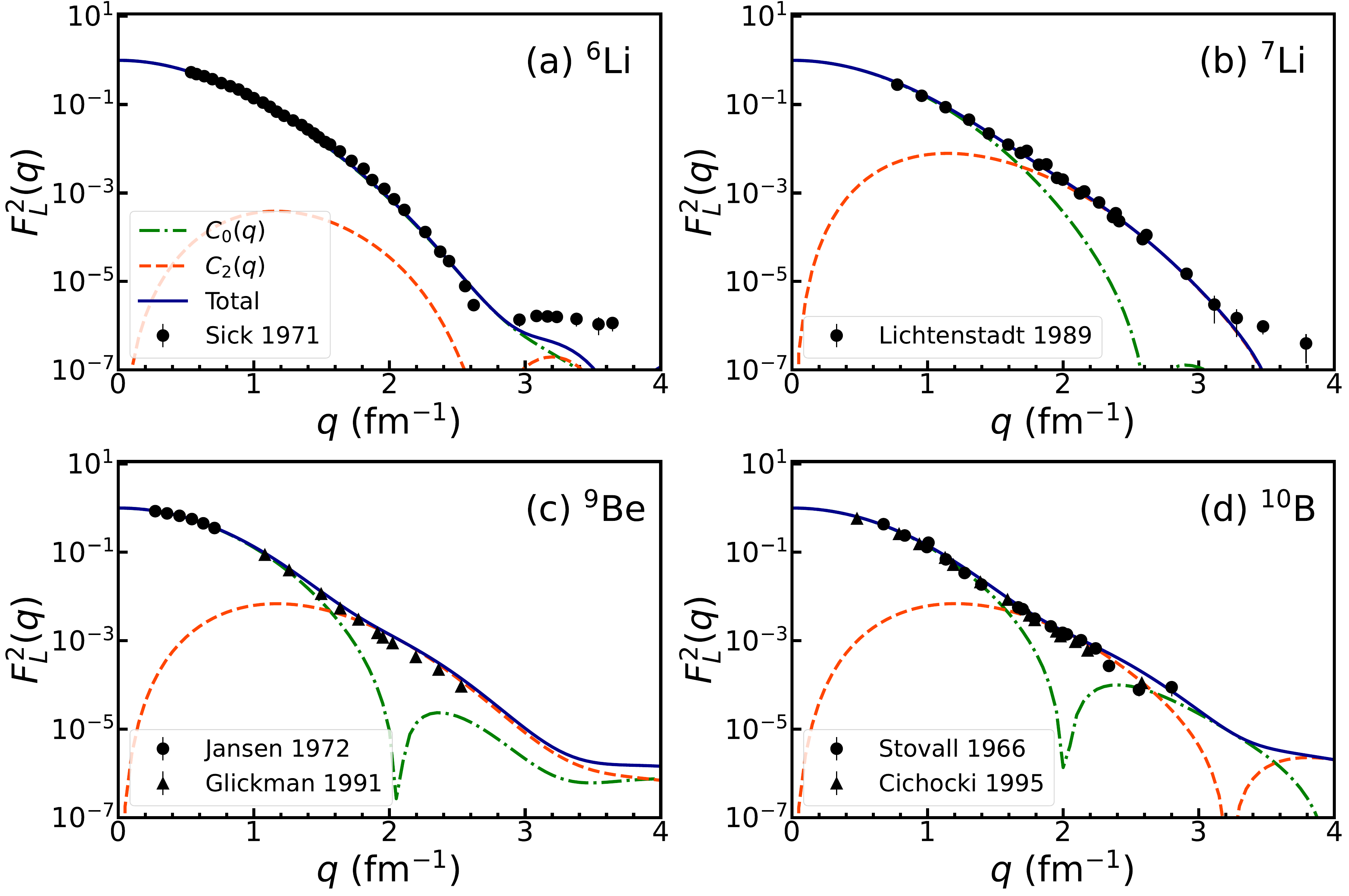}
    \caption{The same as Fig.~\ref{fig:multipole.119} but using model NV2+3-Ia$^{\star}$}  
\label{fig:multipole.116}
\end{figure*}

\subsection{$J\leq 1/2$: $A\,\leq\,4$ isotopes}

We begin our discussion of the results by focusing on the simplest case of $J\leq 1/2$. For these values of angular momenta, only the multipole $C_0(q)$ is at play. 

Let us first focus on the three-body systems $^3$H and $^3$He, which have received significant theoretical attention~\cite{Carlson:1997qn,Bacca:2014tla,Schiavilla:1990zz,Golak:2005iy,Marcucci:2015rca, Piarulli:2012bn,Schiavilla:2018udt,Gnech:2022vwr,Hiyama:2024zfd}. In Fig.~\ref{fig:all.model.he}, we display model calculations using all eight Norfolk models for isotopes of He. Panel (a) displays calculations of $F^2_L(q)$ retaining two-body charge operators through N3LO for $^3$He compared to data from Refs.~\cite{Mccarthy:1977vd,Ottermann:1985km,Amroun:1994qj}. We use blue, orange, green, and magenta lines to denote models I$^{\star}$, I, II$^{\star}$, and II, respectively, while dashed lines represent cutoff ``a" and solid lines represent cutoff ``b". For $^3$He, there is a very clear separation between the cutoffs (``a" vs ``b") and minimal dependence on the energy range (I vs II) or three-body force fit (starred vs un-starred models). While presently we may only estimate uncertainties with this comparative approach, in the future, robust uncertainty quantification of these model dependices should be made possible by using next generation $\chi$EFT potentials fit using sophisticated statistical approaches~\cite{Bub:2024gyz}. The harder cutoff models are better able to reproduce the diffraction minimum in the data, and provide a larger shoulder in the form factor past ${\approx}3.5$ fm$^{-1}$. This is related to the deviation in the two models around the soft cutoff of model ``a" at $\approx 2.5$ fm$^{-1}$, which we observe across nuclei (see, for example, Fig.~\ref{fig:model.comp}). We note, however, that no model is able to fully capture the magnitude of the second peak in the data. The shoulder was not captured in the previous VMC calculation of Ref.~\cite{Schiavilla:1990zz}; however, it was nicely reproduced in the VMC results of Ref.~\cite{Wiringa:1991kp} using a slightly different phenomenological model. Additionally, more recent Faddeev~\cite{Golak:2005iy} calculations using the Argonne $v_{18}$ (AV18)~\cite{Wiringa:1994wb} nucleon-nucleon interaction plus Urbana IX (UIX)~\cite{Pudliner:1995wk} three-body force slightly underestimate the shoulder, while hyperspherical harmonics calculations~\cite{Marcucci:2015rca} using the $\chi$EFT interactions of Ref.~\cite{Entem:2003ft} were unable to reproduce the magnitude of the second peak. 

In Fig.~\ref{fig:model.comp} (a), we show two representative calculations for $^3$H using the NV2+3-Ia$^{\star}$ and NV2+3-IIb$^{\star}$ models. We see that the model with the harder cutoff does a better job of reproducing the shape of the data at larger momentum transfers, similar to the pattern observed for $^3$He. The previous VMC~\cite{Schiavilla:1990zz} and Faddeev~\cite{Golak:2005iy} analyses using phenomenological potentials and currents similarly reproduced the shape of the $^3$H data, while the hyperspherical harmonics calculations using a different $\chi$EFT potential~\cite{Entem:2003ft} and consistent currents under-predicted the location of the diffraction minimum in $q$.

To show the impact of two-body effects, we display calculations for the NV2+3-IIb$^{\star}$ for all isotopes under study using only the LO charge operator and retaining charge contributions through N3LO in Fig.~\ref{fig:two.body} with dashed orange and solid blue lines, respectively. As it has been well established~\cite{Schiavilla:1990zz,Wiringa:1991kp,Golak:2005iy,Marcucci:2015rca,Hiyama:2024zfd}, the role of two-body charge operators on $F^2_L(q)$ is to lower the position of the diffraction minimum in $q$ and to provide more strength in the subsequent peak. In panels (a) and (b), we see this effect clearly for $^3$H and $^3$He, respectively. 

Moving up in mass number, we perform a similar analysis using all Norfolk models compared to data~\cite{Frosch:1967pz,Mccarthy:1977vd,Ottermann:1985km} for $^4$He in Fig.~\ref{fig:all.model.he} (b). While the harder cutoff models still produce diffraction minima lower in $q$ than their soft cutoff counter parts, there is more model dependence arising due to the energy range used to constrain the $N\!N$ interaction and three-body force fits. The harder cutoffs also predict larger strengths in the first shoulder of the form factor; however, no NV2+3 model can capture the magnitude of the data past ${\approx} 3.2$ fm$^{-1}$. Hyperspherical harmonics calculations using the AV18+UIX interaction~\cite{JeffersonLabHallA:2013cus}, as well as VMC and Green's function Monte Carlo calculations using a slightly less sophisticated version of the Argonne plus Urbana model~\cite{Wiringa:1991kp} provided good agreement with both the diffraction minimum and first shoulder in the data. Shell model calculations were also performed in Ref.~\cite{Karataglidis:2006vn} and fair favorably at low $q$, but fail to produce a minimum at $q \approx 3.2$ fm$^{-1}$. Finally, hyperspherical harmonics calculations based on a different set of $\chi$EFT interactions could not provide both an accurate placement of the diffraction minimum and prediction of the magnitude of the shoulder within a given model, but the spread of calculations with different cutoffs is in good agreement with the data in this ${\approx} 3$ to 4 fm$^{-1}$ range~\cite{Marcucci:2015rca}. 
 
It is worth noting that, while for the NV2+3 models a harder cutoff tends to provide a better description of the first diffraction minimum in these systems, the opposite effect was observed in the $\chi$EFT calculations reported in Ref.~\cite{Marcucci:2015rca} using a different family of potentials. For the NV2+3 models, the sensitivity of the first minimum to the central density of the target nucleus~\cite{Hiyama:2024zfd} is interpreted in our calculations as a result of the harder cutoff family tending to produce smaller interior densities~\cite{Piarulli:2022ulk}. In the future, other accurate many-body $\chi$EFT evaluations of this quantity using different families of interactions would be worthwhile to investigate the cutoff sensitivity of the second peak and to investigate where the theory breaks down.

\subsection{$J=1$: $^6$Li}

To reproduce the ground state $F_L^2(q)$ for $^6$Li, one needs to account for two multipole contributions, $C_0(q)$ and $C_2(q)$. The total $F_L^2(q)$ is shown in Fig.~\ref{fig:model.comp} (d), while the role of two-body currents are on display in Fig.~\ref{fig:two.body} (d). Breakdowns of the total curve through N3LO into different multipole components are shown in panels (a) of Figs.~\ref{fig:multipole.119} and Figs.~\ref{fig:multipole.116} for the NV2+3-IIb$^{\star}$ and NV2+3-Ia$^{\star}$, respectively. 

From Fig~\ref{fig:two.body} (d), we see that two-body effects become increasingly important at larger $q$, and have the role of shifting the predicted diffraction minima in the NV2+3-IIb$^{\star}$ to lower $q$. This feature was observed in the VMC evaluation using the AV18+UIX model in Ref.~\cite{Wiringa:1998hr}. Notably, the NV2+3-IIb$^{\star}$ predicts a minimum and reproduces the shoulder in the data, while neither feature is present for the NV2+3-Ia$^{\star}$. Indeed, the shoulder was not reproduced in a calculation using the NV2+3-Ia$^{\star}$ in Ref.~\cite{Gandolfi:2020pbj}. That same study also showed that soft $\chi$EFT models struggled to reproduce this feature. Comparing panels (a) of Figs.~\ref{fig:multipole.119} and Figs.~\ref{fig:multipole.116}, we see that the shoulder in the NV2+3-IIb$^{\star}$ prediction comes entirely from $C_2(q)$, as was found in  the evaluation of Ref.~\cite{Wiringa:1998hr}. Instead, the NV2+3-Ia$^{\star}$ predicts a rather small $C_2(q)$ for the entire range of $q$. It is worth noting that a shell model study including $C_2(q)$ in Ref.~\cite{Karataglidis:1997} was also not able to capture the shoulder behavior, though it compared favorably at low $q$. We conclude that the NV2+3 model differences arise because $C_2(q)$ is sensitive to the particulars of the shape and density profile of the nucleus. Since the NV2+3 models produce different density distributions~\cite{Piarulli:2022ulk}, it is not surprising that they generate different features in $F^2_L(q)$. It could be beneficial to perform calculations using other many-body methods and interactions at $q\gtrsim 2.8$ fm$^{-1}$ to provide constraints on nuclear structure models, due to the model sensitivity of the shoulder. 

\subsection{$J=3/2$: $^7$Li and $^9$Be}

A comparison of $F_L^2(q)$ using the models NV2+3-IIb$^{\star}$ and NV2+3-Ia$^{\star}$ for $^7$ Li is shown in Fig~\ref{fig:model.comp} (e). Compared to the $A\,\leq\,6$ systems, the deviation between the two models is minimal. Two-body current effects, as seen in Fig.~\ref{fig:two.body}, shift the curve to lower $q$ at higher momenta. These effects become sizeable beyond ${\approx} 3$ fm$^{-1}$, changing $F_L^2(q)$ by $\gtrsim 50\%$ in this region. 

Both models agree well with the data from Ref.~\cite{Lichtenstadt:1989nr}. Ref.~\cite{Kanada:1980zz} performed a calculation of $F_L^2(q)$ using the resonating group method and, similarly to our evaluation, found that $C_2(q)$ becomes dominant for $q \gtrsim 2$ fm$^{-1}$, as seen in panels (b) for Figs.~\ref{fig:multipole.119} and~\ref{fig:multipole.116}. In Ref.~\cite{Kanada:1980zz}, this was attributed to the large intrinsic quadrupole deformation of $^7$Li. Good agreement at low $q$ was also obtained in the shell model calculation of Ref.~\cite{Karataglidis:2006vn}, though the data is slightly overpredicted past $q\approx$ 1 fm$^{-1}$. 

The $^9$Be ground state similarly has a strong quadrupole deformation, and we find that $C_2(q)$ is necessary to reproduce the data from Ref.~\cite{Glickman:1991zz} in the range of $1.6~{\rm fm}^{-1} \,\lesssim\, q \,\lesssim\, 2.6~{\rm fm}^{-1}$, which is shown in panels (c) of Figs.~\ref{fig:multipole.119} and~\ref{fig:multipole.116}. The low $q$ data from Refs.~\cite{Glickman:1991zz} and~\cite{Jansen:1972iui} are captured predominantly by $C_0(q)$. 

As in other calculations, two-body currents move the curve to lower $q$ and become more sizeable with increasing $q$. There is a mild model dependence for $^9$Be, seen in Fig.~\ref{fig:model.comp}, with the NV2+3-Ia$^{\star}$ slightly overpredicting the data beginning at $q \approx 1.6$ fm$^{-1}$. The NV2+3-IIb$^{\star}$, instead, begins to slightly underpredict the data starting near $q \approx 2$ fm$^{-1}$. The models also disagree on the shape and magnitude of $C_2(q)$ at larger momentum transfers. The NV2+3-Ia$^{\star}$ predicts that it should provide the dominant contribution up to $q=4$ fm$^{-1}$, while the NV2+3-IIb$^{\star}$ prediction indicates a diffraction minimum in $C_2(q)$ at $q \approx 3.2$ fm$^{-1}$, causing $C_0(q)$ to provide the bulk of the contribution in the range $2.7~{\rm fm}^{-1} \,\lesssim\, q \,\lesssim\, 3.4~{\rm fm}^{-1}$. Because the total curve for the two models differs at large $q$, renewed attention in studying elastic electron scattering on $^9$Be for momentum transfer $q \gtrsim 2.8$ fm$^{-1}$ could help to further constrain models of nuclear structure. 

\subsection{$J=3$: $^{10}$B}

In principle, one may excite four multipoles that contribute to the elastic $F_L^2(q)$ for the $^{10}$B ground state; however, higher-order multipoles are momentum suppressed~\cite{Walecka:1995}. Indeed, in analyzing their experimental data, the authors of Ref.~\cite{Cichocki:1995zz} found that contributions beyond $C_2(q)$ were not needed to explain the data. We thus make the approximation that $C_{l>2}(q)=0$ and solve a system of two linear equations to get the lowest order multipoles. We find that, like the previous experimental analysis concluded, computing only these two multipoles with our many-body framework is sufficient to describe their data and that of Ref.~\cite{Stovall:1966}. The comparison of the NV2+3 calcualtions and data is shown in Fig.~\ref{fig:model.comp} (g). As for the other nuclei under study, two-body currents again shift the curve to lower values of $q$ and are more sizeable as $q$ increases. 

In the region where data is available, both the NV2+3-IIb$^{\star}$ and NV2+3-Ia$^{\star}$ compare favorably and there is very little model dependence. Beyond $q \approx 3$ fm$^{-1}$, the disagreement becomes stark, with model NV2+3-IIb$^{\star}$ predicting an $F^2_L(q)$ that is orders of magnitude smaller than the NV2+3-Ia$^{\star}$. When comparing panels (d) of Figs.~\ref{fig:multipole.116} and Figs~\ref{fig:multipole.119}, both multipole components differ across the models and lead to the enhanced $F_L^2(q)$ for the NV2+3-Ia$^{\star}$. New data for $q \gtrsim 3$ fm$^{-1}$ could not only help further constrain nuclear models by better understanding this deviation, but they could also motivate the computation of higher-order momentum suppressed mulitpoles in the future.

\section{Conclusions}
\label{sec:conclusions}

In this work, we followed upon the studies in Ref.~\cite{Chambers-Wall:2024uhq,Chambers-Wall:2024fha} and computed longitudinal form factors for elastic electron scattering in nuclei with mass $3\,\leq\,A\,\leq\,10$. Our many-body calculations were all performed using the VMC method with models from the NV2+3 family of nuclear interactions. Using the interactions, we saw how different approaches to fitting the potential could-- in some cases-- lead to very different results for the form factors. Further, we were able to assess the contributions for two-body currents. Finally, we carried out the first study of this quantity in the $7\,\leq\,A\,\leq 10$ mass range to use accurate many-body methods based on interacting nucleon degrees of freedom.

Because of the sensitivity of these data to the particulars of the nuclear shape and density, renewed interest in investigating this quantity would help to further constrain nuclear models. In particular, higher $q$ data for $^9$Be and $^{10}$B were identified as particularly impactful for understanding the structure of these systems, as well as validating nuclear Hamiltonians and understanding where the $\chi$EFT framework breaks down. Should new experimental techniques produce elastic longitudinal form factors for unstable isotopes~\cite{Tsukada:2023}, it would provide another window into understanding the structural details of these nuclei. 

This work also opens the doors for several studies in the future. Using the NV2+3 interactions, one could investigate inelastic longitudinal and magnetic form factors in light nuclei. Following the formalism of this work and Refs.~\cite{Chambers-Wall:2024uhq,Chambers-Wall:2024fha}, it is also possible to investigate the role of higher-order many-body processes in both elastic and inelastic electromagnetic form factors. Because the scheme adopted here is not constrained by any particular many-body method, these works could provide templates for the inclusion of higher-order contributions to multipoles involved in studies of heavier systems, as well as a benchmark for other calculations in light nuclei. Finally, with the form factors calculated in our series of papers, we can now compute elastic electron scattering cross sections using a $\chi$EFT approach at different kinematics and compare them with experimental analyses. Work toward such an analysis is currently in progress. 

\acknowledgments 

The authors thank Rocco Schiavilla for his caring mentorship and scientific guidance over the years. 

 This work is supported by the US Department of Energy under Contracts No. DE-SC0021027 (G.~B.~K., G.~C.-W., and S.~P.), DE-AC02-06CH11357 (R.B.W.), DE-AC05-06OR23177 (A.G.), a 2021 Early Career Award number DE-SC0022002 (M.~P.), the FRIB Theory Alliance award DE-SC0013617 (M.~P.), and the NUCLEI SciDAC program (S.P., M.P., and R.B.W.). G.~B.~K. would like to acknowledge support from the U.S. DOE NNSA Stewardship Science Graduate Fellowship under Cooperative Agreement DE-NA0003960. G.~C.-W. acknowledges support from the NSF Graduate Research Fellowship Program under Grant No. DGE-213989. We thank the Nuclear Theory for New Physics Topical Collaboration, supported by the U.S.~Department of Energy under contract DE-SC0023663, for fostering dynamic collaborations.
 A.G. acknowledges the direct support of Nuclear Theory for New Physics Topical collaboration.

The many-body calculations were performed on the parallel computers of the Laboratory Computing Resource Center, Argonne National Laboratory, the computers of the Argonne Leadership Computing Facility (ALCF) via the INCITE grant ``Ab-initio nuclear structure and nuclear reactions'', the 2019/2020 ALCC grant ``Low Energy Neutrino-Nucleus interactions'' for the project NNInteractions, the 2020/2021 ALCC grant ``Chiral Nuclear Interactions from Nuclei to Nucleonic Matter'' for the project ChiralNuc, the 2021/2022 ALCC grant ``Quantum Monte Carlo Calculations of Nuclei up to $^{16}{\rm O}$ and Neutron Matter" for the project \mbox{QMCNuc}, and by the National Energy Research
Scientific Computing Center, a DOE Office of Science User Facility
supported by the Office of Science of the U.S. Department of Energy
under Contract No. DE-AC02-05CH11231 using NERSC award
NP-ERCAP0027147.

\bibliography{biblio}

\appendix

\section{Explicit expressions of the multipoles}
\label{sec:c.of.q}

In this section, we derive the explicit expressions for the reduced charge multipole operators $C^J_l(q)$ for $J=0,~1,~1/2,~3/2,~{\rm and}~3$. For ease of notation, we define,
\begin{equation}
X(q,\cos\theta) =  \mel{JJ}{\rho(q,\theta)}{JJ}\, .
\end{equation}
Recall from Section~\ref{sec:multi} that only even $l$ multipoles will contribute and there will be a finite number of multipoles excited by the scattering; namely, for integer $J$, there are $J+1$ multipoles to compute, while for half-integer $J$, there are $J+1/2$. This means the most trivial cases to consider are $J=0~{\rm and}~1/2$, where only one multipole is excited. We can straightforwardly solve these cases by noting that,
\begin{eqnarray}
c^0_{000} = 1, & c^{\frac{1}{2}}_{\frac{1}{2}\frac{1}{2}0} = \frac{1}{\sqrt{2}} \, ,
\end{eqnarray}
and that $P_0(x) = 1$ for all $x$. Then the only non-zero multipole $C^J_0(q)$ in these cases is given by,
\begin{eqnarray}
C^0_0(q) &=& \frac{1}{\sqrt{4\pi}}X(q,0)\, , \label{eq:c0.j0}\\  
C^{\frac{1}{2}}_0(q) &=& \frac{1}{\sqrt{2\pi}}X(q,0) \label{eq:c0.j1half} \,.
\end{eqnarray}

For $J=1$, we now must solve for two multipole operators. In order to obtain two multipoles, we must make two independent selections of $\theta$. We select $\theta=0$ and $\theta=\pi/4$ and, under this choice, the two equations to invert are given by,
\begin{eqnarray}\nonumber
X(q,0) &=& \sqrt{4\pi}\left[ P_0(0) c^1_{110} C^1_0(q) - \right. \\ 
&& \hspace{1cm} \left. P_2(0)c^1_{112}C^1_2(q) \right] \\ \nonumber
X\left(q,\frac{1}{\sqrt{2}}\right) &=& \sqrt{4\pi}\left[ P_0\left(\frac{1}{\sqrt{2}}\right) c^1_{110} C^1_0(q) - \right. \\ 
&& \hspace{1cm} \left. P_2\left(\frac{1}{\sqrt{2}}\right)c^1_{112}C^1_2(q) \right] \, .
\end{eqnarray} 
When inverted, this system of equations yields,
\begin{eqnarray}
C^1_0(q) &=& \sqrt{\frac{1}{12\pi}} \left[ X(q,0) + 2X\left(q,\frac{1}{\sqrt{2}}\right) \right] \label{eq:c0.j1} \\
C^1_2(q) &=& \sqrt{\frac{8}{3\pi}} \left[ X(q,0) - X\left(q,\frac{1}{\sqrt{2}}\right) \right] \, . \label{eq:c2.j1}
\end{eqnarray}
This is easily extended to $J=3/2$ if we make the substitution,
\begin{eqnarray}
c^1_{110} &\to& c^{\frac{3}{2}}_{\frac{3}{2}\frac{3}{2}0} = \frac{1}{2} \\  
c^1_{112} &\to& c^{\frac{3}{2}}_{\frac{3}{2}\frac{3}{2}2} = \frac{1}{2}\, ,
\end{eqnarray}
which yields,
\begin{eqnarray}
C^{\frac{3}{2}}_0(q) &=& \frac{1}{3\sqrt{\pi}} \left[ X(q,0) + 2X\left(q,\frac{1}{\sqrt{2}}\right) \right] \label{eq:c0.j3half} \\
C^{\frac{3}{2}}_2(q) &=& \frac{4}{3\sqrt{\pi}} \left[ X(q,0) - X\left(q,\frac{1}{\sqrt{2}}\right) \right] \, . \label{eq:c2.j3half}
\end{eqnarray}

In the case of $J=3$, there are now, in principle, four multipoles to solve for and thus one must make four choices of theta. Because higher order multipoles are momentum suppressed~\cite{Walecka:1995} and have been shown as unimportant to describe the low-energy electron scattering data of the $^{10}$B ground state, we can make the approximation that $C_4(q)=C_6(q)=0$ for the $q$ range considered. As shown in~\cite{Cichocki:1995zz}, this is a valid approximation to describe the low-momentum transfer data available for $^{10}$B. Thus, we can again make a simple substitution,
\begin{eqnarray}
c^1_{110} &\to& c^3_{330} = \frac{1}{\sqrt{7}} \\  
c^1_{112} &\to& c^3_{332} = \frac{5}{2\sqrt{21}}\, ,
\end{eqnarray}
resulting in the expressions,
\begin{eqnarray}
C^3_0(q) &=& \frac{1}{6}\sqrt{\frac{7}{\pi}} \left[ X(q,0) + 2X\left(q,\frac{1}{\sqrt{2}}\right) \right] \label{eq:c0.j3} \\
C^3_2(q) &=& \frac{4}{5}\sqrt{\frac{7}{3\pi}} \left[ X(q,0) - X\left(q,\frac{1}{\sqrt{2}}\right) \right] \, . \label{eq:c2.j3}
\end{eqnarray}

\end{document}